# Study of delamination in REBCO coated conductor by transmission electron microscopy

Yan Xin, Jun Lu, and Ke Han

*Abstract*— **Delamination strength of REBCO is very important for its applications in large magnet projects. This work presented the transmission electron microscopy (TEM) investigation of the microstructures of the REBCO coated conductor to understand its delamination property. We found that the low delamination strength is associated with nano-voids formed at the IBAD MgO/Y$_2$O$_3$ interface.**

## I. INTRODUCTION

The second generation high temperature superconductor (HTS) rare-earth-barium-copper-oxide REBCO tapes are excellent conductor for building ultrahigh field superconducting magnets [1], [2]. REBCO are typically grown on ion beam assisted deposited (IBAD) buffer layers on Hastelloy® C-276 polished substrates [3]. There have been a few property issues in its application, among which is the delamination of REBCO's layered structure that degrade the critical currents. The delamination might happen due to the handling during the coil fabrication, due to thermal stress during magnet cool down, and due to the electromagnetic forces during operation [4]. Therefore, the toughness of REBCO against delamination is very important. There are a number of studies transverse mechanical strength using different methods, such as peeling test, the double cantilever beam method, climbing drum test, and anvil mechanical test [5-9]. The location of delamination are found to be inside the REBCO layer, between RBECO layer and the buffer, within the buffer stack, or between Ag/REBCO top surface, and in mixed fashion [10-12].

Understanding why the delamination occurs at particular interfaces and identifying the underlying cause of a weak interface is essential to improve the transverse mechanical strength. There have been very few high resolution microstructures characterization to study the cause of delamination, and identify the delaminating interface. Therefore, in this work, we study the interfaces of REBCO multilayered structure using various transmission electron microscopy (TEM) techniques, with specifical focus on the possible delaminating interfaces. With the detailed microstructure analysis and comparison of REBCO samples of the weak and strong peel strength, we reveal weak interface and the possible causes of it.

## II. EXPERIMENTALS

REBCO samples are made at SuperPower Inc. by the IBAD/MOCVD method grown on Hastelloy C-276 substrate. Its layer structure is depicted in Fig. 1 (not to scale).

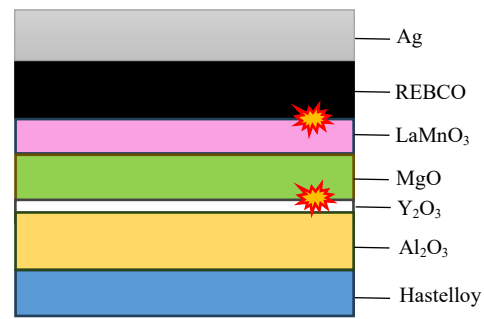

**Fig. 1.** a schematic of layer structure of SuperPower REBCO tape. The possible delamination interfaces are indicated.

The tape is 4 mm wide with 20 μm electroplated copper. As part of the incoming material quality assurance for the 40 T all-superconducting magnet project at the National High Magnetic Field Laboratory, the delamination strength of REBCO are tested for every 100 – 200 m piece length by 90 degree peel strength at room temperature [13]. In this work, we study the interfacial microstructure of a sample that has a peel strength of 1.02 N/cm (strong tape) and a sample with peel strength of only 0.12 N/cm (weak tape).

The cross-sectional TEM samples were prepared by focused ion beam (FIB) on a Helios G4 scanning electron microscope (SEM). The microstructures of the tapes were studied by scanning transmission electron microscopy (STEM) using the probe-aberration-corrected JEOL JEM-ARM200cF at 200 kV. High-angle annular scanning transmission electron microscopy (HAADF-STEM) imaging and the annual-bright-field STEM (ABF-STEM) imaging were the two major imaging techniques used. Electron energy loss spectroscopy (EELS) and energy dispersive X-ray spectroscopy (EDS) spectrum imaging (SI) were used in STEM mode with a probe size of 0.11 nm using Gatan GIF and Oxford Aztec EDS detector.

This work was performed at the National High Magnetic Field Laboratory, which is supported by the National Science Foundation Cooperative Agreement No. DMR-1644779, DMR-1839796, DMR- 2131790, and the State of Florida. (*Corresponding author: Yan Xin and Jun Lu.*)

Yan Xin, Jun Lu, and Ke Han are with Magnet Science and Technology division at National High Magnetic Laboratory, Tallahassee, FL 32310 USA (e-mail: xin@magnet.fsu.edu; junlu@magnet.fsu.edu; han@magnet.fsu.edu ).

Color versions of one or more of the figures in this article are available online at http://ieeexplore.ieee.org

## III. RESULTS

### A. The surface of the weak tape after delamination

After the 90 degree peel test, it was observed that both sides of the delaminated strong tape showed black color. This indicates that delamination occurred within the REBCO layer which is black. In contrast, the weak tape delaminated within the buffer stack manifested by the yellowish color on the substrate side of the delaminated tape. The yellowish color comes from the interference of the lights reflected from different interfaces of the buffer stack which is transparent.

As shown in Fig. 2(a), the weak tape was peeled leaving the superconductor layer (black) attached to the copper side. The substrate side showing yellow color for most area, with small patches of teal blue. To determine the delaminated surfaces of both yellow and blue areas, we used FIB to cut a TEM lamella that includes both regions as indicated in the inset of Fig. 2 (b). The SEM cross-sectional view of the TEM sample is shown in Fig. 2 (b) indicate more buffer layers remain on the blue color area than the yellow color area. The elemental EDS maps (Fig. 2 (c) to (h)) revealed that the surface for the yellow area is $Y_2O_3$ and the blue area is $LaMnO_3$ (LMO). The cross-sectional STEM view of the intersection between yellow and blue areas is shown in Fig. 2 (j). The further magnified image Fig. 2(k) indicates reasonable bonding between MgO lattice planes and those of the $Y_2O_3$ grains, which suggest that the $MgO/Y_2O_3$ interface at this boundary is relatively strong, and the weaker link is at the REBCO/LMO interface instead. This is why the delamination occurred at REBCO/LMO interface at this point, leaving a patch of LMO layer (blue) at the surface. However, Fig. 2 (j) also reveals nanometer sized voids at $MgO/Y_2O_3$ interface. The detailed microanalysis of these nano voids will be presented in section III C.

### B. The REBCO/$LaMnO_3$ interface

The results shown in Fig. 2 suggests that both REBCO/LMO and $MgO/Y_2O_3$ interfaces are possible delamination locations. To compare the REBCO/LMO interface of the strong and weak tapes, a cross-sectional TEM sample was cut in width direction for each of the as-received tapes. Fig. 3(a) and (b) are ABF-STEM images that show REBCO/LMO interfaces of the strong and the weak tape respectively. The darker contrast at interface of the weak tape (Fig. 3(b)) suggests interfacial strain. The HAADF-STEM images in Fig. 3(c) (strong tape) and 3(e) (weak tape) are not significantly different. Both show good epitaxial relationships of REBCO/LMO with the atomic planes across the REBCO/LMO interface. Further analysis is performed by inverse FFT where REBCO(006)/LMO(101) spots, indicated by red circles in the inset of Fig. 3 (d) and (f), are used to obtain the inverse FFT images (Fig. 3 (d) and (f)). They show that the strong tape has coherent interface, whereas the weak tape has two extra atomic planes as labelled by the red lines and red arrows in Fig. 3(f). The extra atomic planes is indicative of interfacial edge dislocations which might be one of the reasons for delamination at REBCO/LMO interface.

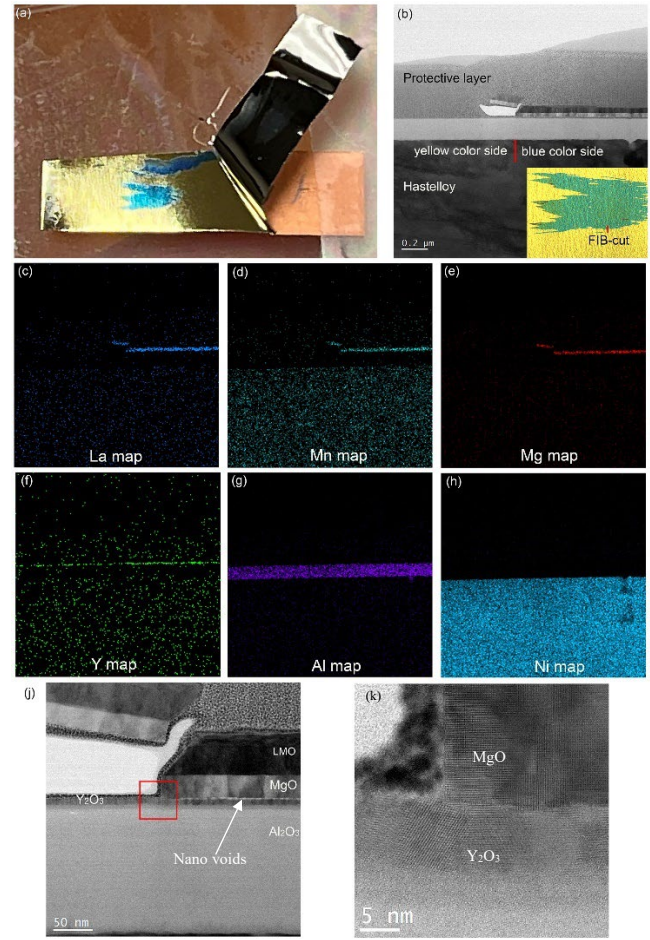

**Fig. 2.** (a) a picture of the delaminated weak tape. (b) ABF-STEM image of the cross-sectional view at the yellow/blue boundary. Inset: a picture of the blue patch. (c) to (h) elemental EDS maps. (j) and (k) high resolution ABF-STEM image of cross-sectional view of the joint at the yellow/blue boundary. (k) higher magnification image of the red box in (j)

### C. $MgO/Y_2O_3$ interface

Based on the observations presented in section III A, the delamination of the weak tape mostly occurs at the $MgO/Y_2O_3$ interface. Therefore, we examine this interface of both samples for comparison. The $MgO/Y_2O_3$ interface of both strong and weak tapes as shown in Fig. 4. Compared to the $MgO/Y_2O_3$ interface of the strong tape (Fig.4(a)), this interface of the weak tape shows a contrast of a bright band in the TEM bright field (BF) image (Fig.4(b))

To accurately image the interfaces without delocalization effect of BF-TEM, ABF-STEM was used, and results are shown in Fig. 4(c) and 4(d). Compared to the smaller and more separated bright area at the $MgO/Y_2O_3$ interface of the strong tape (Fig. 4(c)), the band in the weak tape is wider and brighter (Fig. 4(d)) indicating less material at the interface. The corresponding HAADF-STEM images (Fig. 4(e) and 4(f)) also show darker contrast at this interface. It is well known that the intensity of HAADF-STEM image is proportional to $Z^{1.7}$ (Z is atomic number of the element) as well as sample thickness along the electron beam direction [14]. Since atomic number of Y is much greater than Mg and Al, $Y_2O_3$ layer shows brighter contrast than MgO and $Al_2O_3$ layer. The dark contrast at the interface corresponds to nano-voids which results in less thickness along the electron beam direction. Apparently, the interfacial nano-voids are larger and closer together in the weak tape (Fig. 4(f)). The average dimension of the nano-voids is about 3 nm x 2 nm for the strong tape, and 5 nm x 2.5 nm for the weak tape. Their dimensions in the depth direction (perpendicular to the paper) will be discussed below in junction with Fig. 5.

On closer examination of the $MgO/Y_2O_3$ interface of the weak sample (Fig. 5(a) and (b)), it is discernable that the $Y_2O_3$ layer is comprised of three regions: R1, nano-void region of about 2 nm just below the MgO layer; R2, the crystalline $Y_2O_3$ region of about 7 nm; and R3, the amorphous region of 4 nm of $Y_xAl_yO$ which is the intermixture of $Y_2O_3$ and the $Al_2O_3$ layer below. The strong tape has similar microstructural morphology except the R1 region. Along the $MgO/Y_2O_3$ interface at different location, the nano-voids appears to have different shape and sizes, as shown in Fig. 5(c) and (d).

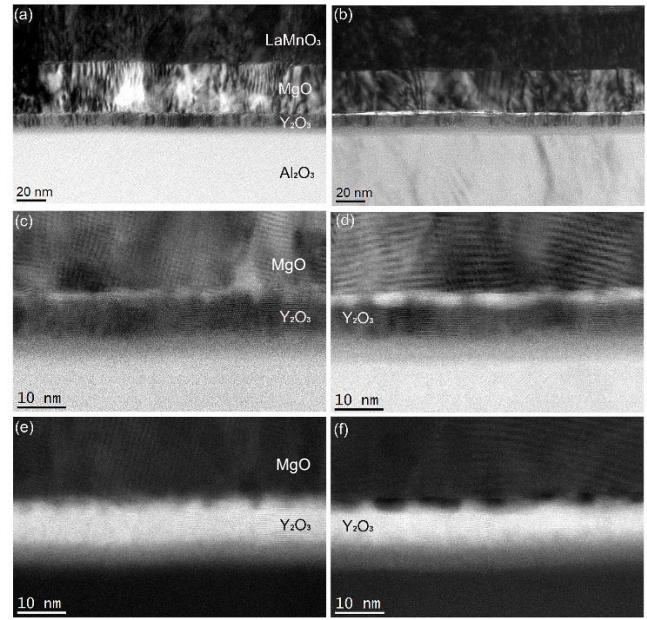

**Fig. 4.** General cross-sectional view of the $MgO/Y_2O_3$ interface of both strong and weak tape. (a) TEM BF image of the strong tape. (b) TEM BF image of the weak tape. (c)-(d) ABF-STEM images of the strong tape. (e)-(f) HAADF-STEM images of the weak tape.

In order to measure the depth of the nano-voids, we collected the EELS spectra from the interfacial region of both samples. and used EELS log-ratio method to obtain the thickness map of this region. Fig. 5(e) shows a region (red box) of the weak sample where EELS spectra are taken. Fig. 5(f) the thickness map (by EEL-spectra imaging) corresponding to the red box region. The measured depths of $Y_2O_3$ at these voids are $20 - 25$ nm out of total of 38 nm. Therefore, the size of the void in depth direction is $13 - 15$ nm (about $34 - 39\%$ of total). It should be noted that this 13-15 nm corresponds to one or more voids in the depth direction. In comparison, the depth of voids in the strong tape is only 7 nm out of total of 65 nm (11%).

It is important to verify whether the interfacial voids are results of contamination in the deposition process. The elemental EDS maps of the weak tape are shown in Fig. 6 where no contaminating elements, typically carbon, are segregated at the $MgO/Y_2O_3$ interface, although the $Y_2O_3$ layer has a small amount of carbon distributed through the entire layer. Moreover, the EDS spectrum taken from the interface region (the box in Fig 6(a)) is shown in Fig. 6 (g). Again, no significant amount of contaminating was detected except Ar which seem to exist in the entire $Y_2O_3$ layer.

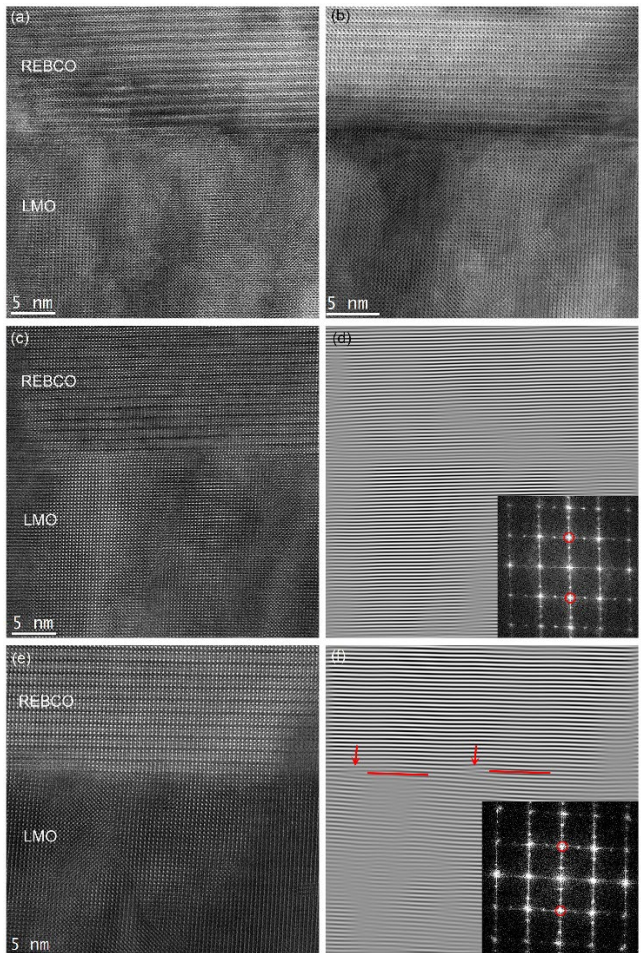

**Fig. 3.** (a) ABF-STEM image of the REBCO/LMO interface of a strong tape. (b) ABF-STEM image of the REBCO/LMO interface of a weak tape. (c) Corresponding HAADF-STEM image of (a). (d) Inverse FFT of (c). inset: FFT diffraction pattern. (e) Corresponding HAADF-STEM image of (b). (f) Inverse FFT of (e). inset: FFT diffraction pattern.

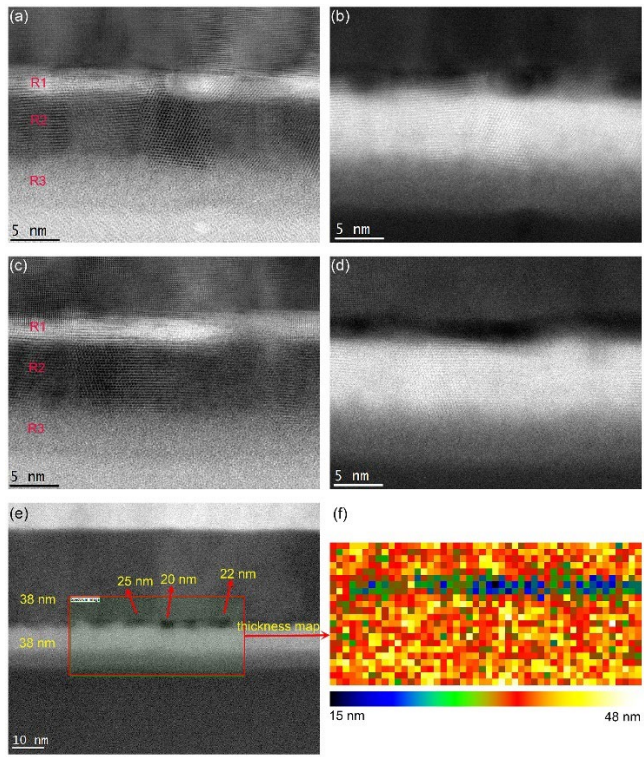

**Fig. 5.** Atomic resolution images of the MgO/Y$_2$O$_3$ interfaces of the weak tape. (a) and (b) a pair of ABF-STEM and HAADF-STEM images of one region of the interface. (c) and (d) a pair of ABF-STEM and HAADF-STEM images of another region of the interface. (e) The interface region where the EEL-Spectra Image was collected. (f) The thickness map of the region.

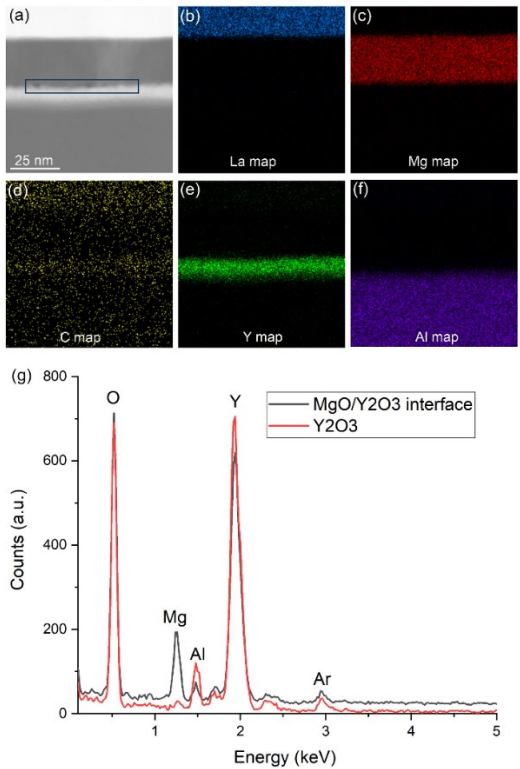

**Fig. 6.** Elemental EDS maps: (a)STEM image of mapped region; (b) La map; (c) Mg map; (d) C map; (e) Y map; (f) Al map. (g) EDS spectra from the MgO/Y$_2$O$_3$ interface (black) and from the middle of Y$_2$O$_3$ layer (red).

## IV. DISCUSSIONS

Our microscopy analysis clearly shows that the high density nano-voids at the MgO/Y$_2$O$_3$ interface is responsible for the low delamination strength of the weak sample. We have noticed that these nano-voids are present in both samples, and not uniformly distributed. The higher fraction of nano-voids at the interface, the weaker delamination strength of the tape.

We speculate that these nano-voids originated during the IBAD MgO growth. In the IBAD process, Ar ion beam bombards the Y$_2$O$_3$ surface while the MgO is growing by magnetron sputtering or by thermal evaporation. When the parameters of the Ar ion beam deviates from its optimal conditions, it might result in a relatively rough MgO/Y$_2$O$_3$ interface. During the subsequent REBCO growth at higher temperature, the amorphous Y$_2$O$_3$ layer turns into crystalline, turning the rough interface into nano-voids at the interface.

Our investigation (not presented here) also shows that the crystal quality of the superconductor layer of both strong and weak tapes are similar. Consequently, their superconducting properties are similar, as confirmed by their critical current (Ic) measured at 77 K self-field. Therefore, it seems that nano-voids at the MgO/Y$_2$O$_3$ interface of the weak sample do not compromise the crystal quality and the electrical performance of REBCO.

## V. CONCLUSION

Delamination strength of REBCO is very important for its applications in large magnet projects. To identify the layer location of the delamination, we studied the interfaces of SuperPower REBCO tapes that show strong and weak peel strength by transmission electron microscopy techniques. We found nano-voids at the IBAD MgO/Y$_2$O$_3$ interface which is likely the cause of the weak peel strength. Dislocations at the REBCO/LaMnO$_3$ interface added unwanted stress and might have contributed to weakening REBCO/LMO interface.